# How the Fusion of Onboard Sensors and V2X Data can improve (or not) the Cooperative Perception of Connected Automated Vehicles

Amir Mohammadisarab, Miguel Sepulcre, Luca Lusvarghi, Javier Gozalvez
Uwicore lab., Universidad Miguel Hernández de Elche (UMH), Spain.
Email: {amohammadisarab, msepulcre, llusvarghi, j.gozalvez}@umh.es

***Abstract*— Automated vehicles rely on onboard sensors to perceive their surroundings and navigate autonomously. However, sensor performance may degrade under adverse weather conditions or when line-of-sight is obstructed. Cooperative perception (or collective perception) is expected to mitigate these limitations by enabling Connected and Automated Vehicles (CAVs) to share sensor data and collaboratively enhance situational awareness. Several studies have analyzed the potential of cooperative perception, yet the fusion of V2X data with information from onboard sensors has received limited focus. V2X data may contain errors that affect the quality of the fused data, and hence the effectiveness of cooperative perception. This study analyzes the impact of sensing measurement errors, V2X packet losses, and GNSS inaccuracies on the effectiveness of cooperative perception. The results highlight the potential of cooperative perception to enhance perception levels and range compared to using onboard sensors alone. However, they also identify key challenges related to the generation of ghost vehicles during the fusion process, which must be addressed to prevent V2X data from introducing additional errors when fused with onboard sensor data.**

## I. INTRODUCTION

Cooperative perception (or collective perception, CP) is essential for improving the perception of connected and automated vehicles (CAVs) [1]. An automated vehicle relying solely on onboard sensors faces two main challenges in perceiving its environment: the inherent limitations of its field of view and sensing range, and the presence of obstructions that can block the sensors' line-of-sight. Through CP, CAVs can exchange perception data about detected objects (including kinematic and geometric information) via vehicle-to-everything (V2X) communication thereby extending their field of view and sensing range. To do so, data from V2X cooperative perception must be fused with locally acquired onboard sensor data [2]. This process is typically decomposed into three main subprocesses: temporal and spatial alignment, association, and fusion. Due to variations in positioning systems across different vehicles and delays in V2X communication, the received information about detected objects must first be spatially and temporally aligned with onboard sensor data. Next, association algorithms determine which detected objects from different sources (onboard sensors and other CAVs) correspond to the same real-world object. Finally, a fusion algorithm integrates these common observations into a unified representation [2].

The fusion of V2X and onboard data can be performed at various levels, including low, medium, and high levels [3]. Low-level fusion integrates raw sensor data from different sources, while medium-level fusion combines extracted features. High-level fusion operates at the object or track level, where each sensor independently performs filtering and tracking before fusing the outputs from all sensors. While fusing raw data or features provides finer details, it strains the limited available spectrum. Consequently, current ETSI [4] and SAE [5] standards for cooperative perception have opted for a high-level fusion architecture, as CAVs exchange object-level data via V2X.

Existing research on multi-sensor high-level fusion for autonomous vehicles (e.g. [3],[6]) has demonstrated the importance of effective track-to-track (or object-level) fusion algorithms. However, research on the fusion of V2X cooperative perception data with onboard sensor data is still in its early stages. Studies on the design of scalable cooperative perception often assume a simplified fusion process, in which the fused object set is merely the union of objects detected by onboard sensors and those received in cooperative perception messages [1][7][8]. Therefore, these studies cannot accurately quantify the impact of inaccuracies or errors in V2X data on the effectiveness of cooperative perception once V2X data is fused with onboard sensors data at the ego vehicle. Robust and consistent methods for the spatial and temporal alignment of information before track-to-track fusion have been proposed [9], demonstrating that errors can be reduced even in the presence of V2X communication latency [10]. Other studies analyze the fusion of cooperative perception data at the infrastructure level, but without considering its fusion with onboard sensors [11]. Despite existing advancements, the impact of different noise sources on the accuracy of the fusion between V2X and onboard data remains underexplored. Existing studies often assume ideal sensor measurements, reliable V2X data exchange, and perfect GNSS positioning. In contrast, this paper analyzes the impact of sensing noise, packet losses, and GNSS errors on the quality of V2X and onboard data fusion, and consequently, on the effectiveness and challenges of cooperative perception.

## II. HIGH-LEVEL DATA FUSION

In this paper we adopt the high-level sensor data fusion architecture from [3] and expand it for the fusion of onboard sensor data with V2X data received through cooperative perception. The architecture is shown in Fig. 1. It consists of two levels: 1) sensor and 2) fusion. We assume that track-to-track fusion occurs at time step $t$, and omit the time index in all notations for simplicity.

This work has been partly funded by MCIN/AEI/10.13039/501100011033 and the "European Union NextGenerationEU/PRTR" (TED2021-130436B-I00).

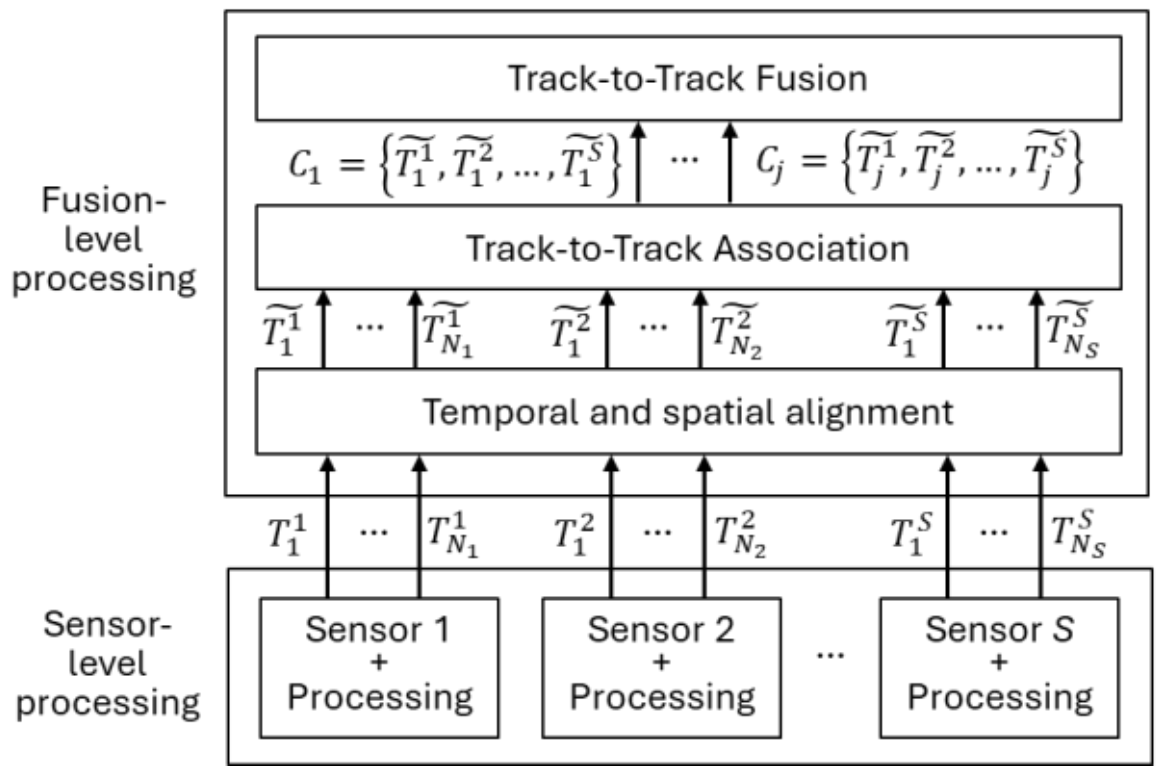

Fig. 1. High-level data fusion architecture for onboard sensors.

### A. Sensor-level processing

In a high-level fusion architecture, each sensor observes the environment and processes the raw data independently of the other sensors. This sensor-level processing essentially includes feature extraction, classification and tracking [3]. Although the sensors work independently, all of them must produce an output that is compatible with the next level of the architecture, i.e., the fusion-level processing. The output of each sensor, denoted as sensor $j$, is a track list (also referred to as an object list) represented by $T^j = \{T_1^j, \dots, T_{N_j}^j\}$, where $N_j$ is the number of objects detected by sensor $j$. A track $T_i^j$ contains at least the state vector and covariance matrix [3][12] of detected object $i$ perceived by sensor $j$, and is denoted as $T_i^j = \{X_i^j, P_i^j\}$. The state vector $X_i^j$ can include the object position, speed, acceleration, heading, etc. The covariance matrix $P_i^j$ represents the uncertainty in the estimation of the state and is provided by the tracking algorithm (e.g. a Kalman filter). Specifically, the covariance matrix quantifies the statistical relationships (variances and covariances) between the different elements of the state vector being estimated. The application of a Kalman filter on a sequence of position and speed measurements can smooth out noise in the data, providing more accurate and consistent estimates of the detected object's state.

### B. Fusion-level processing

All the tracks generated by the sensor level are used as inputs for the fusion level. At the fusion level, the tracks must first be aligned temporally and spatially. This is because each track may be produced using a different time reference and coordinate system (e.g., each sensor might be located in a different position on the vehicle). As illustrated in Fig. 1, track $T_i^j$ of object $i$ obtained by sensor $j$ is transformed to $\widetilde{T_i^j}$ after the temporal and spatial alignment.

The next step in the fusion level is the track-to-track association. The goal of the association is to group the tracks representing the same object in a cluster. Each cluster is identified by a unique ID. Ideally, each cluster represents one object, and therefore the cluster ID can be considered equivalent to the object ID. In Fig. 1, the cluster of object $i$ is denoted as $C_i$. Hungarian, Auction, or Greedy algorithms can be used to solve the association as an assignment problem to minimize or maximize a cost function (e.g., distance).

Finally, the track-to-track fusion is performed to combine the tracks belonging to the same cluster into a unified and consistent track. The goal is to improve accuracy, robustness, and coverage by leveraging the strengths of each sensor. The covariance matrix can be used to provide more influence or priority to the tracks with higher accuracy. Simple methods could be employed for the track-to-track fusion, such as Weighted Average Fusion that computes a weighted average of the state estimates (e.g., position or velocity) from the different tracks using weights inversely proportional to the uncertainty of each track. More advanced methods include the Covariance Intersection (CI) [12], where the fused estimate is computed using a weighted combination of means and covariances. CI is especially useful when track correlation is unknown. The number of outputs of the track-to-track fusion is equal to the number of clusters, which is ideally equal to the number of objects detected by the sensors.

Fig. 2 illustrates in a 2D data plane, the ground truth positions of three objects, four tracks obtained from four sensors, their association, and the position resulting from the track-to-track fusion process.

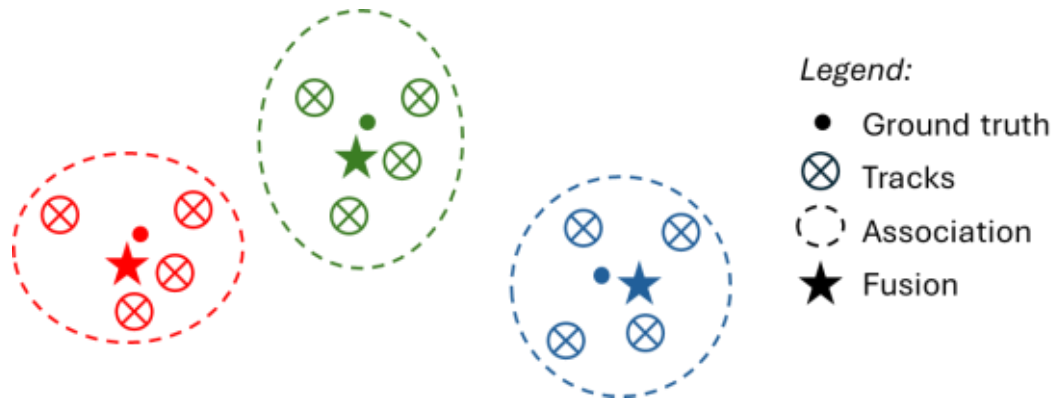

Fig. 2. Ground truth, tracks, association and fusion in a 2D data plane.

## III. V2X DATA FUSION

We leverage the high-level sensor data fusion architecture in Fig. 1 to fuse V2X data with onboard sensors data. For this purpose, we consider that each CAV transmitting cooperative perception messages acts as a virtual sensor at the sensor-level processing (Fig. 1) for the ego vehicle. The number of virtual sensors is then equal to the number of cooperative perception messages received at time $t$. This approach requires that the tracks produced by the onboard sensors and the V2X data have the same format at the sensor-level processing. This is a reasonable assumption because cooperative perception messages include information such as the position and speed of the object, the measurement time, as well as the correlation matrix for selected components of the provided kinematic state (including position and speed). A key difference between the V2X data and sensor data is that the V2X data may result from the fusion process performed by the transmitting CAV. CAVs may also intentionally filter the transmitted data based on its quality and omit some objects according to object inclusion rules. The reception of V2X messages may also fail due to propagation errors or packet collisions.

We adopt a sensor-to-sensor fusion strategy, also referred to as memoryless fusion [3][13], at the fusion-level processing. To isolate the effects of different noise sources on the fusion pipeline, we assume an ideal temporal and spatial alignment following [10]. This assumption implies that all sensor measurements are generated synchronously at both the ego vehicle and the transmitting CAVs. The track-to-track association algorithm considered in this study is an adaptation of the Greedy algorithm proposed in [13]. The Greedy approach evaluates pairwise Euclidean distances between tracks and iteratively assigns them to clusters based on the smallest distances, ensuring that tracks from the same sensor do not share a cluster. The adapted Greedy approach introduces a merging step for tracks originating from different sensors to better handle scenarios involving multiple sensors. The track-to-track fusion algorithm adopted is covariance intersection [12]. It is considered a robust method when the correlations between tracks are unknown, which is particularly useful for fusing onboard sensor tracks with those transmitted by CAVs through cooperative perception messages. The covariance intersection method combines the state vectors and the covariance matrices from multiple sensors into a single fused track while ensuring consistency. To achieve this, it uses a weighted combination of the inverse covariances, with the weights optimized to minimize the fused uncertainty.

## IV. EVALUATION SCENARIO

We analyze the impact of sensing measurement errors, V2X packet losses, and GNSS inaccuracies on the effectiveness of cooperative perception on a 5 km highway with 4 lanes (2 lanes per direction). The mobility of the vehicles is generated with SUMO (Simulation of Urban MObility) at a traffic density of 60 veh/km. This study assumes that half of the vehicles in the scenario are CAVs (50% penetration rate) and that all CAVs transmit cooperative perception messages. We assume a 50% penetration rate of CAVs to reflect a plausible near-future deployment scenario where both connected and non-connected vehicles coexist. In each simulation, the ego vehicle and the CAVs are selected randomly. The ego vehicle is chosen from the two central lanes and the central portion of the highway (from 1.5 km to 3.5 km), while CAVs can be located anywhere within the highway scenario. To ensure statistically robust results, we conduct over 1000 simulations, with each simulation lasting 10 seconds.

Vehicles are equipped with *S*=5 sensors providing a sensing range of 200 meters and a field of view of 360º in unobstructed environments. An object is detected by a sensor if it is located inside its sensing area and is not obstructed by other objects (i.e., other vehicles in our scenario). Each sensor provides a new measurement and track per detected object every 100 ms, and all sensors are synchronized. All sensors are capable of measuring the position and speed of the detected objects, albeit with certain distance-dependent noise[14]. T The noise is modeled with a zero-mean Gaussian distribution that is added to the ground truth position and velocity of each detected object. The standard deviation of this sensing noise is determined based on the distance between the object and the sensor ($d_{o,s}$). We evaluate both a low and a high sensing noise model. The low sensing noise model considers the same standard deviation for the positioning noise and velocity as in [10]:

$$\sigma_{low}^{pos}(d_{o,s}) = \sigma_{low}^{vel}(d_{o,s}) = d_{o,s}/1000 \quad (1)$$

The high sensing noise model considers the same velocity error, but a larger positioning error following:

$$\sigma_{high}^{pos}(d_{o,s}) = 0.01 + d_{o,s}/20 \quad (2)$$

$$\sigma_{high}^{vel}(d_{o,s}) = d_{o,s}/1000 \quad (3)$$

In addition, we assume that each sensor employs a Kalman filter to track the objects it detects and to generate the corresponding covariance matrices. If a sensor detects an object for the first time, no tracks are generated; the first measurement is used to initialize the Kalman filter for that particular object.

This work assumes that all CAVs transmit a cooperative perception message every 100 ms that includes the relative position, speed, and covariance matrices of all detected objects. The inclusion of information about all detected objects is currently adopted by SAE [5] and is also supported for implementation in [4] (using the *ObjectInclusionConfig* flag). When the ego vehicle receives a cooperative perception message, it must transform the relative position of the objects included therein to their absolute position using the GNSS position of the transmitter as a reference. Thus, the GNSS error represents an additional source of noise, which we model with a zero-mean Gaussian distribution having a standard deviation of 2 m [15]. CAVs transmit cooperative perception messages using C-V2X Mode 4 with a 23 dBm transmission power, MCS 9 (QPSK modulation and a coding rate of approximately 0.7), and 4 sub-channels per sub-frame. Packet losses are modeled with the analytical model proposed in [16], which considers half-duplex errors, signal sensing errors, propagation errors, and errors due to packet collisions.

## V. RESULTS

We evaluate the performance in six different scenarios. In the first two scenarios, cooperative perception is not active and the ego vehicle detects objects using only its onboard sensors. Detection can be performed without (scenario O1) or with (scenario O2) sensor measurement errors. In the four other scenarios (from V1 to V4), cooperative perception is active and the ego vehicle uses onboard sensor data and V2X data to detect objects. To assess the effectiveness of fusing V2X and onboard data in the presence of noise sources, we incrementally introduce each noise source to identify their individual effects. Scenario V1 represents an ideal case with no sensing errors, no V2X packet losses, and no GNSS errors at the ego vehicle or the CAVs. In scenario V2, sensing errors (low or high) are present in the measurement of the position and speed of the detected objects. Scenario V3 adds V2X packet errors following the analytical model proposed in [16], and scenario V4 introduces GNSS errors, thereby representing the scenario with all potential sources of error. Table I summarizes the six different scenarios.

TABLE I. SCENARIOS

(a) Onboard sensor only

| ID | Sensing error |
|---|---|
| O1 | No |
| O2 | Yes |

(b) Onboard sensors + V2X

| ID | Sensing error | Packet loss | GNSS error |
|---|---|---|---|
| V1 | No | No | No |
| V2 | Yes | No | No |
| V3 | Yes | Yes | No |
| V4 | Yes | Yes | Yes |

Fig. 3 plots the OAR (Object Awareness Ratio) as a function of the distance between the ego vehicle and the detected objects. The OAR metric is defined to quantify how accurately the fusion process detects existing objects (vehicles in our scenario). We use the global nearest neighbor (GNN) algorithm to determine which detected object at the output of the fusion pipeline corresponds to which ground truth object. GNN is based on the minimum distance between detected and ground truth objects, with the constraint of one-to-one matching. In this study, we compute the OAR as the total number of detected–ground truth object pairs with a localization error of less than 5 meters, divided by the total number of ground truth objects.

Without cooperative perception (scenarios O1 and O2), Fig. 3 shows that the OAR is limited to approximately 71% at short distances (0 to 50 meters) and declines as the distance from the ego vehicle increases. The same trend is observed for both low (Fig. 3a) and high (Fig. 3b) sensing noise. The sensors of the ego vehicle are unable to correctly detect all surrounding objects because they are often blocked by other vehicles. Since the probability that an object is blocked increases with distance, the OAR degrades accordingly. Fig. 3a demonstrates that, without cooperative perception, the presence of low sensing noise has a negligible effect on the ability to detect existing objects. This is because low sensing noise minimally impacts the localization error of detected objects with respect to their actual location, and therefore on the OAR. This is visible in Fig. 4a, which plots the localization error of detected objects before fusion-level processing using boxplots. In contrast, high sensing noise significantly increases the localization error (Fig. 4b), thereby impacting the OAR (Fig. 3b). Fig. 3b shows that the OAR without cooperative perception is reduced from 71% to 50% at short distances when high sensing noise is present compared to the ideal case without any noise. Large localization errors can cause the association algorithm to fail to correctly group the tracks of an object for fusion, which degrades the OAR. For example, it can mistakenly associate tracks belonging to different objects or create ghost objects that do not really exist. These effects are illustrated in Fig. 5.

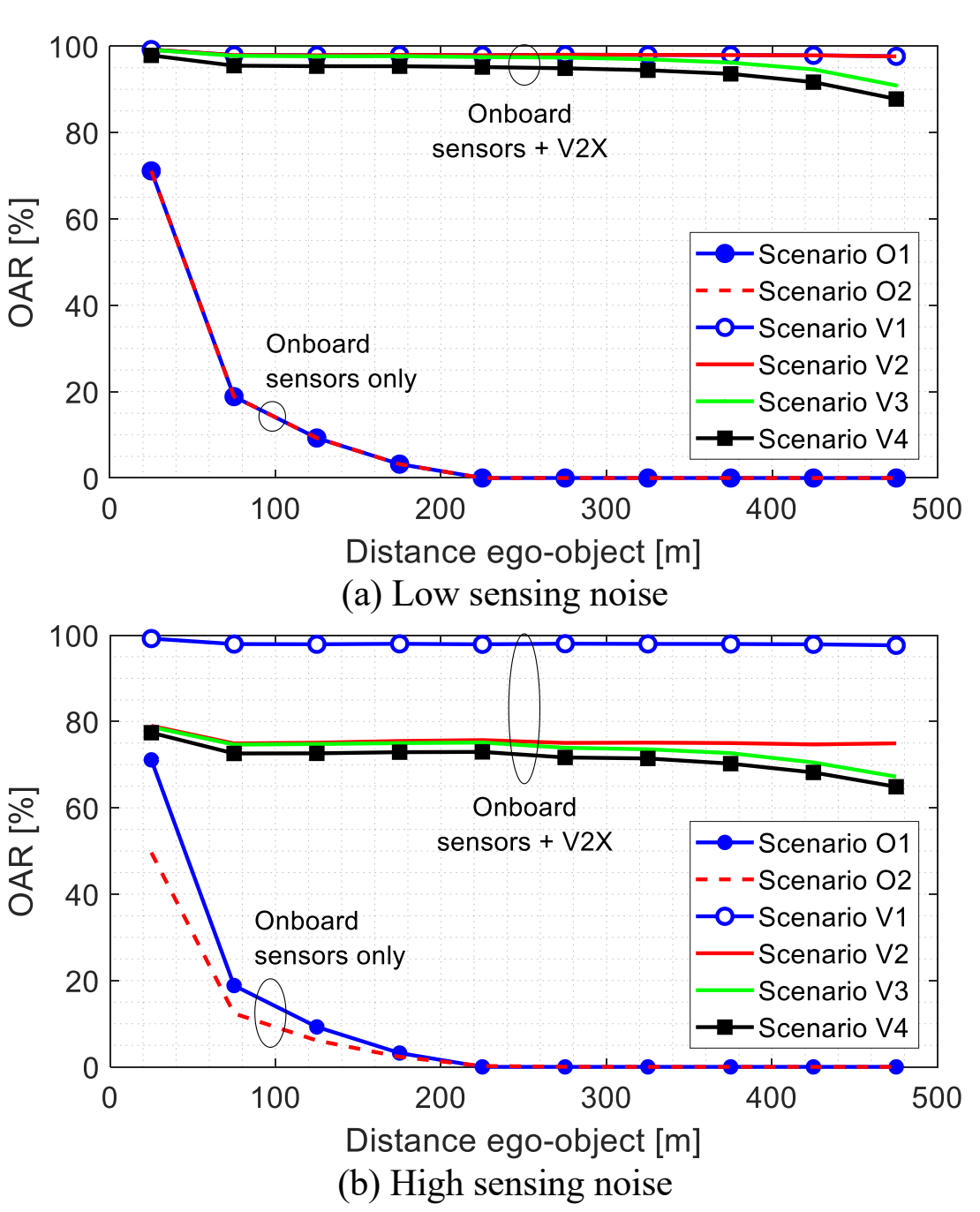

(a) Low sensing noise

(b) High sensing noise

Fig. 3. OAR (Object Awareness Ratio) as a function of the distance between the ego vehicle and the object.

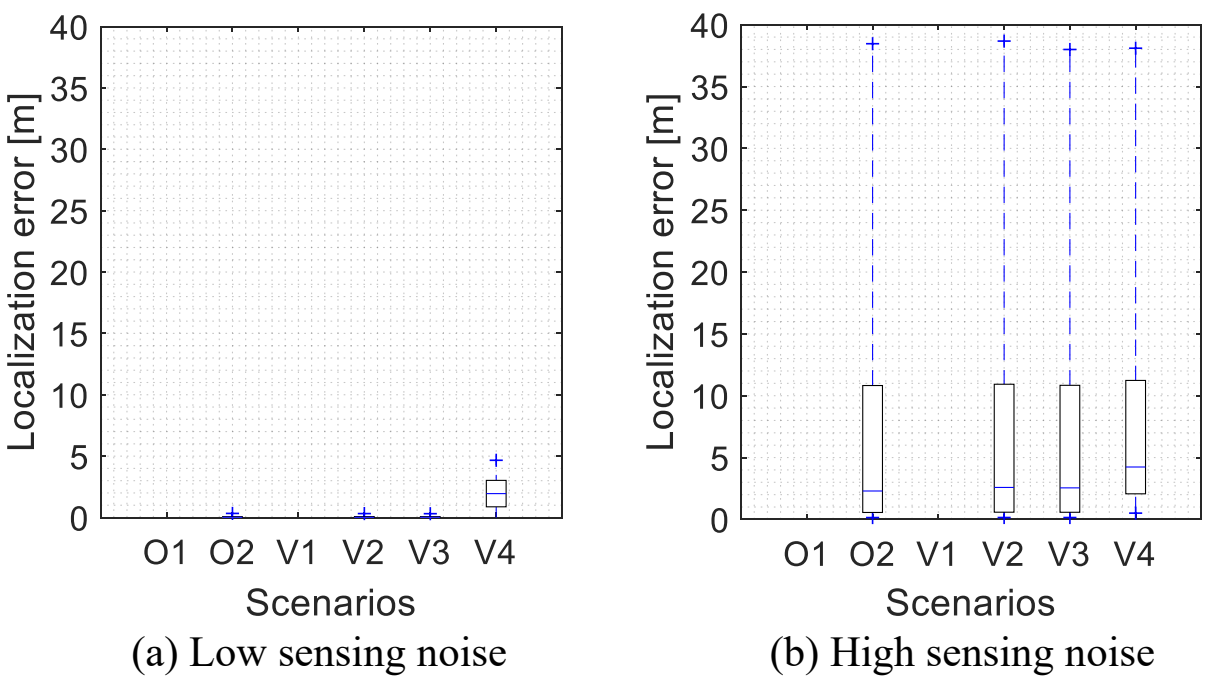

(a) Low sensing noise (b) High sensing noise

Fig. 4. Localization error before the fusion level processing for low and high sensing noise. Each box represents the median and 25th and 75th percentiles, and the whiskers show the 5th and 95th percentiles.

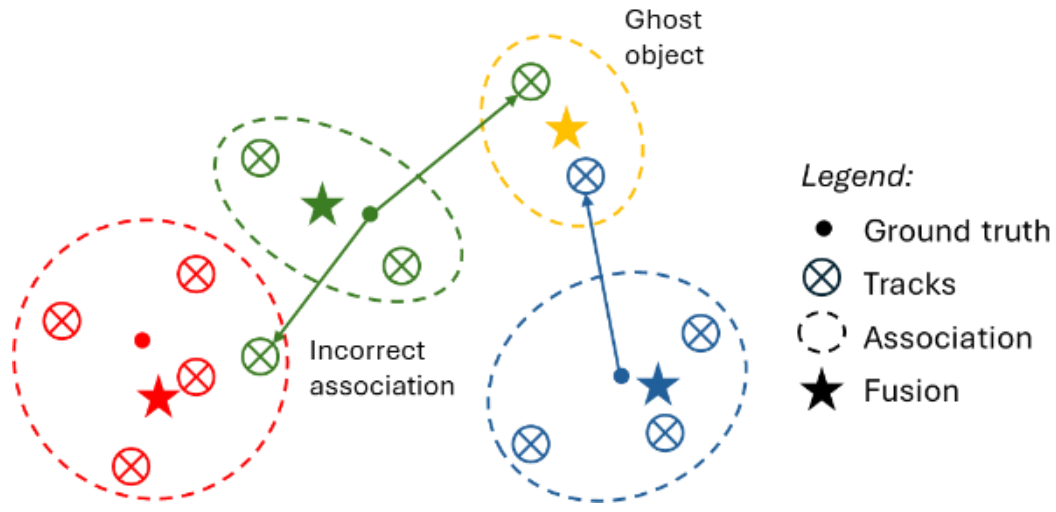

Fig. 5. Ground truth, tracks, association and fusion in a 2D data plane with noisy sensing measurements that provoke incorrect associations and ghost objects.

Fig. 3 also shows the OAR obtained with cooperative perception, i.e., when onboard sensor data is fused with the received V2X data. Without any errors (scenario V1), the OAR remains around 98%, irrespective of the distance to the ego vehicle. This indicates that the ego vehicle receives accurate information about nearly all objects via V2X. The remaining 2% correspond to vehicles that are not detected by any CAV and thus not included in any cooperative perception message. The OAR does not degrade with low sensing noise in scenario V2 (Fig. 3a) since the localization error is very low (Fig. 4a). However, with high sensing noise, the localization error increases (Fig. 4b), and the OAR achieved with cooperative perception is significantly reduced from 98% to around 75% (Fig. 3b).

The introduction of packet losses in scenario V3 maintains the OAR at short and medium distances but degrades it at large distances, as shown in Fig. 3. Comparing scenario V3 with V2, packet losses produce a similar degradation of the OAR in both low and high sensing noise scenarios (Fig. 3a and Fig. 3b), with approximately an 8% reduction in OAR at 500 m. The degradation in PDR (Packet Delivery Ratio) at large distances (e.g. 35% at 500m) is compensated by the redundancy in object detection by several CAVs—on average, around 3.4 CAVs detect the same object in the scenario and report it via V2X. Thus, the probability that at least one cooperative perception message containing information about the same object is correctly received by the ego vehicle remains quite high even at large distances. In fact, the average number of messages received by the ego vehicle containing information about the same object from different CAVs is 2.8 when packet losses are modeled. Fig. 3 also illustrates the effect of GNSS error in scenario V4. Compared to scenario V3, the results show a degradation of the OAR by around 3% due to increased localization error. The effect of GNSS error on localization error is more apparent for low sensing noise (V4 in Fig. 4a). For high sensing noise, the localization error is already high due to the sensing noise, and GNSS error contributes only a relatively small additional increase (V4 in Fig. 4b).

The results in Fig. 3 clearly demonstrate the potential of V2X-based cooperative perception and how CAVs can significantly improve their perception compared to using onboard sensors alone. The improvement in the OAR metric with V2X is mostly due to the fact that more information about more objects is available at the ego vehicle via V2X. This effect is obviously dominant for objects beyond the sensing range of onboard sensors but is also relevant at shorter distances due to occlusions. However, there is also the risk that V2X data can contaminate onboard sensor data. This is evident in Fig. 6, which plots the average number of clusters produced by the association process in each ego-object distance range, which corresponds to the number of detected objects after fusion-level processing. The width of each distance range is 50 m. This metric is computed by counting the number of clusters located in each distance range and by averaging across all considered distance ranges. In the case of onboard sensors only, the maximum ego-object distance range is 150-200 m. When onboard sensors and V2X data fusion is employed, the maximum considered ego-object distance is 450-500 m. Fig. 6 compares this number with the actual number of objects measured in the ideal scenario where cooperative perception is used and no sensing noise, V2X packet losses, or GNSS error is present (horizontal dashed line in the figure). The results are presented in Fig. 6 for both low and high sensing noise scenarios. The figure shows that when only onboard sensors are used, the number of detected objects is lower than the count with cooperative perception in the ideal scenario. This is due to sensor limitations and obstructions that prevent onboard sensors from detecting many objects in the vicinity of the ego vehicle. When V2X data is fused in the low noise scenario (Fig. 6a), the number of detected objects closely approximates the ideal count. However, this number increases under the influence of GNSS error (scenario V4), resulting in the detection of ghost vehicles, i.e., false detections of objects that do not exist. On average, the ego vehicle can detect one more object when GNSS error is present than in the ideal scenario without any error sources. This effect is significantly more pronounced with high sensing noise (Fig. 6b), where a substantial number of ghost objects are created (on average, four) when V2X data is fused with onboard data. This arises because the received tracks are noisy, causing the association algorithm to fail in grouping them correctly. It artificially generates a high number of non-existent objects, as illustrated in Fig. 5. The results in Fig. 6 demonstrate that when V2X data is affected by any type of noise—even solely GNSS error—it can result in the detection of ghost vehicles. Ghost vehicles may confuse the ego vehicle and influence its driving decisions. These findings underscore the need for improved error mitigation and robust association techniques to better manage the inherent uncertainties in V2X data during the fusion process with onboard sensor data.

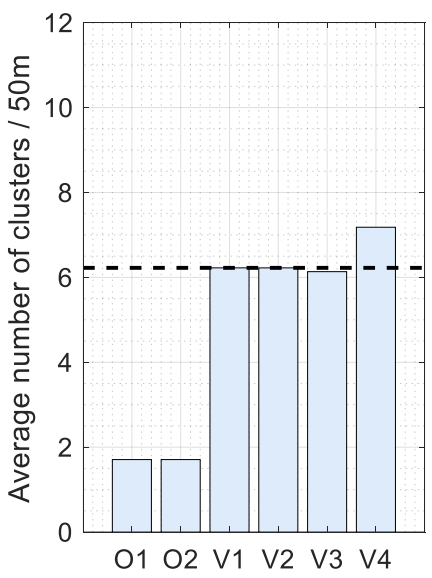

(a) Low sensing noise

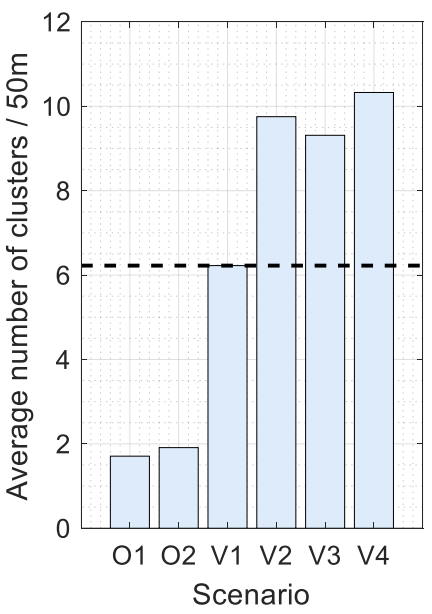

(b) High sensing noise

Fig. 6. Average number of clusters (or equivalently detected objects) every 50m distance to the ego vehicle for low and high sensing noise.

## VI. Conclusions

This study has investigated the high-level fusion of onboard sensor data with V2X data for cooperative perception in connected automated driving. Our analysis focused on the effects of key noise sources—namely, sensing noise, packet losses, and GNSS errors—on overall fusion performance and perception levels. The results demonstrate the high potential of cooperative perception and the fusion of onboard and V2X data under low sensing noise conditions, achieving extended and high perception levels. However, our study has shown that the presence of GNSS errors can artificially create ghost objects that do not truly exist due to errors in the association process. This effect is significantly exacerbated when sensing noise increases. Under high sensing noise, the perception level achieved with the fusion of onboard and V2X data can significantly decrease, although it remains substantially higher compared to using onboard sensor data alone. Nonetheless, high sensing noise conditions amplify the risk of generating ghost objects, which may adversely affect driving decisions. These findings highlight the challenges that must be overcome to ensure that the fusion of onboard and V2X data consistently enhances sensor perception without introducing additional errors. Future work should focus on designing more robust association algorithms and developing strategies to mitigate noise sources, thereby ensuring that only high-quality and useful V2X data is transmitted.